\documentclass[acmsmall]{acmart}
\usepackage{caption}
\usepackage{subcaption}
\usepackage{array}
\usepackage{multirow, booktabs}
\AtBeginDocument{%
  \providecommand\BibTeX{{%
    \normalfont B\kern-0.5em{\scshape i\kern-0.25em b}\kern-0.8em\TeX}}}

\setcopyright{acmlicensed}
\copyrightyear{2024}
\acmYear{2024}
\acmDOI{abcde.XXXXXXX}

 \acmConference[Conference acronym 'XX]{Make sure to enter the correct conference title from your rights confirmation emai}{June,2024}{xx}
\acmISBN{xxxxxx}

\begin{document}

\title{The Decoy Dilemma in Online Medical Information Evaluation}
\subtitle{A Comparative Study of Credibility Assessments by LLM and Human Judges}
\author{Jiqun Liu}
\authornote{Corresponding author.}
\email{jiqunliu@ou.edu}
\orcid{0000-0003-3643-2182}
\affiliation{%
  \institution{The University of Oklahoma}
  \streetaddress{401 W Brooks Street}
  \city{Norman}
 \state{OK}
  \country{USA}
 \postcode{73019}
}

\author{Jiangen He}
\affiliation{%
  \institution{The University of Tennessee, Knoxville}
  \streetaddress{1345 Circle Park Drive}
  \city{Knoxville}
  \state{TN}
  \country{USA}}
\email{jiangen@utk.edu}

\renewcommand{\shortauthors}{Anonymous}

\begin{abstract}
Can AI be \textit{cognitively biased} in automated information judgment tasks? Despite recent progresses in measuring and mitigating social and algorithmic biases in AI and large language models (LLMs), it is not clear to what extent LLMs behave "rationally", or if they are also vulnerable to human cognitive bias triggers. To address this open problem, our study, consisting of a crowdsourcing user experiment and a LLM-enabled simulation experiment, compared the credibility assessments by LLM and human judges under potential decoy effects in an information retrieval (IR) setting, and empirically examined the extent to which LLMs are cognitively biased in COVID-19 medical (mis)information assessment tasks compared to traditional human assessors as a baseline. The results, collected from a between-subject user experiment and a LLM-enabled replicate experiment, demonstrate that 1) Larger and more recent LLMs tend to show a higher level of consistency and accuracy in distinguishing credible information from misinformation. However, they are more likely to give higher ratings for misinformation due to the presence of a more salient, decoy misinformation result; 2) While decoy effect occurred in both human and LLM assessments, the effect is more prevalent across different conditions and topics in LLM judgments compared to human credibility ratings. In contrast to the generally assumed "rationality" of AI tools, our study empirically confirms the \textit{cognitive bias risks} embedded in LLM agents, evaluates the decoy impact on LLMs against human credibility assessments, and thereby highlights the complexity and importance of debiasing AI agents and developing psychology-informed AI audit techniques and policies for automated judgment tasks and beyond.



\end{abstract}



\maketitle

\section{Introduction}
Credibility assessment has been a central theme in information seeking and retrieval (IS\&R) and Machine Learning (ML) research, and plays a key role in online information evaluation and decision-making activities~\cite{yamamoto2011enhancing, kattenbeck2019understanding, bink2022featured, flanagin2020making}. Facilitating and enhancing information credibility assessments is essential for improving individuals' information literacy, informal learning, and overall wellbeing~\cite{zhang2023design, jung2024tech}.  \textit{Medical information evaluation} as a critical information-intensive task contextualizing credibility assessments faces significant challenges, such as social and cognitive biases~\cite{liu2023behavioral, liu2024search}, online misinformation and disinformation~\cite{ioannidis2017survive, swire2020public}, as well as model hallucination in Artificial Intelligence (AI) systems~\cite{vicario2019polarization, ji2023survey}. In IS\&R and DL experiments, obtaining reliable assessment labels is also challenging due to human biases and heuristics, high cost and inconsistency in user judgments, and variations in tasks and topics~\cite{liu2022toward}. With these obstacles, it is often difficult to evaluate the credibility aspect of IR systems in a scalable manner and reuse human-generated evaluation resources. While a series of recent experiments demonstrate the potential of large language models (LLMs) in judging document quality and predicting user preferences~\cite[e.g.][]{thomas2024large, rahmani2024synthetic, zhang2024large, ma2024leveraging}, it is not clear 1) to what extent LLMs are cognitive biased (e.g. due to the biases inherited from human-generated contents and actions) in a sequence of document judgments, and 2) how well LLMs perform, compared to human judges, in assessing information credibility and detecting medical misinformation and disinformation, especially under the influence of varying \textit{bias triggers} in prompts and training data. Answers to these questions will be instrumental for assessing the feasibility and performance of applying LLMs in online information evaluation and decision-making support, and facilitating unbiased human-AI collaboration in information-intensive tasks. 

As one of the initial steps toward answering the open questions above and exploring potential cognitive biases of LLMs in credibility assessment tasks, our study focuses on \textit{Decoy Effect}~\citep{wedell1996using, wu2020profiting}, a cognitive bias that has been widely discussed and empirically confirmed by cognitive psychology experiments in a variety of domains. Decoy effect, contradicting the predictions of \textit{expected utility} theory and (over)simplified rational models~\citep{mankiw2014principles}, refers to the phenomenon in which the presence of an inferior option can influence an individual's preferences between two similar options within a decision-making task~\citep{thaler2016behavioral}. While recent research has touched upon the potential decoy effect in relevance judgments and ranking, the impact of decoy information on online information evaluation and the judgments of AI agents still remain understudied in computing research~\citep{liu2023toward, chen2023decoy, chen2024decoy}. Our experiments, aiming to address this open problem, are situated within the context of medical information judgment and misinformation detection, one of the critical information tasks that significantly affect public health, healthcare, and individuals' wellbeing~\cite{swire2020public, southwell2019misinformation}. We selected \textit{Web search} as the simulated interaction context in our research for evaluating both human and AI judgments, as it serves as the main channel for individual users to access online medical information and often exposes users to bias triggers and misinformation generated by human and machine agents~\cite{sun2019consumer, soroya2021information, liu2020investigating, chen2023reference, azzopardi2021cognitive}. However, it is worth noting that findings from our research on investigating the credibility judgments by AI and human assessors can potentially be applied beyond search domains (e.g. online discussion forum and social media, news recommendation) and inform the design of auditing and debiasing techniques for LLMs in automated judgment tasks. 

If human assessors can be cognitively biased due to decoy effects, do LLM agents offer us fairer, more rational judgment labels? Or, do LLM agents make mistakes and biased judgments like human assessors do? To address these open questions, we first conducted a \textit{crowdsourcing between-subject user experiment} where we collected participants' credibility assessments on COVID treatment Web pages under \textit{decoy} and \textit{non-decoy conditions}. The labels from human assessors were used as a baseline for measuring the level of biases embedded in LLMs' judgments. Then, we ran a \textit{LLM-based evaluation experiment} where we replicated the experimental condition adopted in user experiment and compared the credibility assessments from several widely-applied LLMs. Findings obtained from human and machine experiments allow us to reveal the extent to which different LLMs are vulnerable to decoy effects compared to human assessors, and to empirically measure LLMs' levels of \textit{rationality} in judgment tasks. Our contributions are threefold:
\begin{itemize}
    \item The comparative study shows that in contrast of the widely-accepted assumptions regarding the rationality of AI, LLMs' credibility judgments have a higher level of decoy-related bias compared to human judgments in a variety of medical topics and assessment contexts. 
    \item Methodologically, our \textit{human-LLM dual experiment setup} can inform the design of future bias mitigation techniques in LLM agents and facilitate the development of human-in-the-loop LLM evaluation benchmarks. 
    \item Regarding practical applications, our study reveals the hidden cognitive biases in LLMs' responses and judgments, and highlights the importance of bias mitigation and psychology-informed AI audit in automated judgments.    
\end{itemize}

\section{Related Work}
This section discusses the theories and research progress from the areas related to our study, which will clarify where our interdisciplinary research contributions are situated in the field. 
\subsection{Cognitive Bias and Decoy Effect}
The decoy effect, also termed the \textit{asymmetric dominance} effect, has been widely examined and is well-recognized in the disciplines of cognitive psychology~\citep{pettibone2000examining, wedell1996using}. The concept was initially introduced by \cite{tversky1985framing} through controlled experiments and characterizes the phenomenon that the inclusion of a decoy option can significantly influence the perceived attractiveness of the available choices, despite no changes to the actual options.  For example, consider a consumer choosing between two smartphones: "\textbf{A}: a basic model priced at \$300" and "\textbf{B}: a premium model priced at \$600." The decision might be difficult to predict. However, when a third option, "\textbf{C}: a mid-range model priced at \$599," is introduced as a decoy, option \textbf{B} may become more appealing, even though the decoy option \textbf{C} is not intended to be chosen. The presence of the mid-range model makes the premium model appear to be a better deal in comparison.

Building on \cite{tversky1985framing}'s early work, subsequent research has delved into the decoy effect and sought to validate its influence across various domains, topics, and decision-making scenarios, such as consumer purchasing decisions~\citep{huber1982adding, wu2020profiting, yuan2022retailer}, medical treatment choices~\citep{stoffel2019testing, blumenthal2015cognitive}, and children's behavior~\citep{zhen2016development}. Beyond observable behavioral changes, \cite{hu2014neural} investigated the neural correlates associated with the decoy effect, and found that choice sets with decoys activated the occipital gyrus while deactivating the inferior parietal gyrus. One possible explanation for the decoy effect is the concept of \textit{salience}, or the extent to which an option stands out. The presence of a decoy option as a \textit{reference point} can shift the relative salience of the original ones, and increase the \textit{perceived attractiveness} of one option~\citep{connolly2013regret, simonson1989choice}. Chen et al.~\cite{chen2023decoy, chen2024decoy} examined the impact of decoy results on document relevance judgments and demonstrated the presence of decoy effects in IR evaluation. Evidence confirming the decoy effect contradicts the "context-invariant" assumption and offers an alternative explanation for seemingly irrational choices. Meanwhile, investigating decoy effects, along with other related cognitive biases, may offer us a more realistic, psychology-informed basis for characterizing and evaluating the process of human judgments. 

In contrast to the controlled and simplified decision-making tasks often employed in behavioral experiments, evaluating medical information in web searches is a much more complex, context-dependent process. This evaluation is shaped by various contextual factors, including the rank position of search results, the nature of the topics being explored, and users' pre-existing beliefs and biases about the information they encounter~\citep{sbaffi2017trust, zhang2015quality, chen2023decoy}. These factors significantly influence how users assess the credibility and relevance of medical information, making them more vulnerable to cognitive biases like the decoy effect. The decoy effect, which skews user preferences by introducing an inferior but comparable option, can distort judgments in high-stakes areas such as healthcare, where misinformation can have serious consequences. Studying this bias in the context of medical information evaluation is crucial, as it can provide key insights into how both search engines and LLMs influence user decision-making, particularly when credibility assessments are involved.

As LLMs are increasingly used for tasks that may rely more on automated judgments~\cite{thirunavukarasu2023large}, especially in high-stake domains (e.g. medical information evaluation, healthcare and finance), understanding the impact of cognitive biases such as the decoy effect on these systems becomes essential. Exploring how this bias manifests in real-world settings will allow us to better assess the value and risks associated with deploying LLMs for such tasks. However, this line of research necessitates the careful design of experiments that can disentangle the mixed effects of multiple contextual variables—such as search result ranking, user intent, and pre-existing beliefs—that shape the decision-making process. Future experimental designs must integrate realistic search scenarios, diverse user profiles, and sophisticated metrics to accurately capture the complex dynamics of user interaction with both search results and LLM-generated content. By conducting such studies, researchers can deepen their understanding of the cognitive mechanisms behind user judgments and develop systems that are better equipped to mitigate the risks posed by cognitive biases, thereby improving trust in AI-enabled systems and building sustainable human-AI collaborations.


\subsection{Information Retrieval Evaluation}
Evaluation has been a central them in IR research, and understanding user assessments and preferences has always been a cornerstone of IR evaluation experiments~\citep{harman2011information, kelly2009methods, hofmann2016online}. Among the various types of judgments, \textit{relevance judgment}—assessing how topically relevant a document is to a given query—has traditionally served as the foundation for decades of system-centered IR evaluation~\citep{voorhees2019evolution, jarvelin2017ir, saracevic2007relevance}. The offline evaluation metrics, based on \textit{query-document relevance}, have set the benchmarks for differentiating high-performing retrieval systems from less effective ones and have been critical in developing scalable and reproducible evaluation of ranking algorithms~\citep{kim2022alignment}.


While the relevance-based Cranfield paradigm has been fundamental in advancing \textit{ad hoc} factual retrieval techniques~\citep{voorhees2019evolution}, its limitations are increasingly apparent when applied to the complexities of real-world information-seeking tasks~\citep{liu2022toward, belkin2008relevance}. The paradigm's emphasis on isolated, relevance-based judgments is insufficient for capturing the nuanced, iterative nature of modern search behavior, particularly in multi-query tasks aimed at resolving complex or exploratory information needs~\citep{belkin2016people}. Real-world information retrieval is rarely confined to a single query-response cycle; rather, it involves users continuously refining queries and synthesizing information across multiple documents as their understanding evolves. Topical relevance alone does not account for the broader range of judgments users make, which are influenced by factors such as information quality, credibility, and trust. Consequently, there is a growing recognition within the IR community of the need to extend evaluation frameworks beyond relevance and to incorporate a more comprehensive view of information-seeking behavior, particularly through the lens of \textit{sessions}, where users engage in multi-query interactions over time~\citep{liu2021deconstructing}. A session-based evaluation approach offers a more realistic and holistic view of user interaction with IR systems, capturing not only the relevance of individual results but also the temporal and cognitive processes that underpin users' judgments and decision-making in more complex search scenarios.

Recent research has increasingly focused on expanding the scope of IR evaluation by examining dimensions beyond topical relevance, with an emphasis on user-centered factors that are critical in real-world contexts. Key dimensions such as \textit{usefulness}~\citep{liu2022toward, cole2009usefulness}, \textit{trustworthiness}~\cite{wang2024trustworthy, polley2021towards, zhang2022learning}, and \textit{credibility}~\citep{yamamoto2011enhancing, wu2020credibility, hilligoss2008developing, lim2013college} are essential in shaping user judgments, particularly in domains where information quality and reliability are essential. In health and medical information seeking and retrieval, for instance, credibility judgments are crucial as users navigate conflicting information from potentially unreliable sources~\cite{wang2021online, sun2019consumer}. These dimensions are often inadequately captured by traditional relevance-based metrics, which highlights the need for more robust evaluation frameworks that account for the broader context in which information is judged. Additionally, retrospective session satisfaction~\citep{al2010review, zhang2020models} has gained prominence as an indicator of how well IR systems meet users' evolving needs across an entire session, offering a richer understanding of system effectiveness beyond isolated interactions. The increasing significance of these user-centered dimensions is amplified by the growing prevalence of misinformation and the integration of AI-generated content into information retrieval systems~\citep[e.g.][]{popat2017truth, pennycook2019fighting, zrnec2022users, vicario2019polarization, ji2023survey}. The rise of LLMs in search platforms has heightened concerns about biases, model hallucination and the amplification of false information, making the evaluation of credibility, trustworthiness, and overall user satisfaction more urgent than ever. Developing IR systems that not only retrieve relevant content but also ensure the quality and trustworthiness of information is vital, particularly in critical domains such as healthcare, science, and public policy. 

Despite these advancements, for IR evaluation tasks, most studies have been conducted in \textit{ad hoc} retrieval settings~\cite[e.g.][]{chen2017meta} or have focused on isolated document judgments without considering the broader search process~\cite[e.g.][]{scholer2013effect, kelly2015many}. \cite{eickhoff2018cognitive} investigated biases, including the decoy effect, in crowdsourcing tasks and their impact on user judgments of documents. Nevertheless, such studies may not able to take into consideration the full context of search session dynamics and are over-restricted by controlled settings~\cite{wang2024cognitively}. Overcoming this limitation requires creating appropriately controlled environments that can trigger the behavioral effects of cognitive biases while maintaining a degree of authenticity in simulated sessions.

\subsection{LLM in Information Judgment Tasks}

The integration of LLMs into IR evaluation frameworks offers significant potential to enrich and advance the field by simulating more sophisticated and behaviorally realistic search interactions~\cite{thomas2024large, salemi2024evaluating}. Unlike traditional interaction models, which often rely on simplified or formalized representations of user behavior, LLMs possess the ability to process and generate natural language in ways that closely mimic human communication and decision-making processes. This capability allows for the creation of richer, more authentic synthetic data that better reflects the complexity of real-world search scenarios~\cite{wang2023improving, sekulic2024analysing}. By leveraging LLMs, researchers can explore a wider range of user behaviors, such as query reformulation, information synthesis, and the negotiation of conflicting information, all of which are essential for understanding user judgments in dynamic and multifaceted information environments. Furthermore, LLMs have the potential to enhance the simulation of multi-query sessions, enabling a deeper examination of how users assess the credibility and relevance of information over time. This is particularly relevant for complex IR tasks, where users may need to evaluate contradictory sources or synthesize information from diverse domains~\cite{abbasiantaeb2024let}. By offering a more nuanced understanding of credibility assessments and other critical dimensions of user interaction, LLMs can contribute to the development of more comprehensive evaluation frameworks that capture the full breadth of user judgments and experiences in IR systems~\cite{sekulic2024analysing, zhai2024large}. These advancements may ultimately lead to more effective, trustworthy IR systems capable of meeting the demands of increasingly complex and information-rich environments.

Driven by recent research progresses in Natural Language Processing (NLP), LLM has been increasingly applied and assessed in tasks that are traditionally completed by human assessors, including information judgment tasks in IR and recommender systems evaluation~\cite{faggioli2023perspectives, zheng2023judging}. The adoption of LLM in these tasks are usually motivated by LLMs' flexibility in prompting, scalability of automated labeling, and the capability in processing customized natural language inputs and outputs~\cite{ji2023survey, thomas2024large, wang2023improving}. Despite the strengths of LLMs, challenges remain in the application of LLMs for information judgments, especially in IR settings~\citep{dai2024bias}. Issues such as model bias and the need for domain-specific training continue to be significant concerns. For example, Chen et al.~\cite{chen2024humans} addressed the bias in LLM outputs and proposed methods to measure fairness and reliability in model predictions. Chen et al.~\cite{chen2024ai} investigated the threshold priming effect in LLM's batch relevance judgment, which may lead to systematic biases in LLM-generated relevance scores and thereby influence the training and evaluation of rankers. In addition, the computational resources required for training and deploying LLMs pose practical limitations~\cite{li2023unified}. Efforts to mitigate these challenges, such as developing more efficient models and refining prompt-tuning and retrieval-augmented generation (RAG) techniques could better adapt LLMs to specific judgment tasks. In addition to the challenges above, the potential \textit{cognitive biases} in judgments, which could result from algorithmic design and human-generated training data, still remains understudied~\cite{liu2023toward, chen2023decoy, lewis2020retrieval, liu2024search}, especially under the default assumptions regarding model and AI rationality in varying fields. 

\section{Research Questions}
With the research challenges and open problems identified above, our study on investigating the credibility assessments from both LLM and human judges seeks to answer two research questions: 
\begin{enumerate}
    \item[\textbf{RQ1}]: To what extent are LLM agents vulnerable to decoy effects compared to human judges in evaluating the credibility of online medical information in Web search?
     \item[\textbf{RQ2}]: To what extent do LLM agents' vulnerability to decoy effects vary across different topics and assessment contexts?   
\end{enumerate}


\section{Methodology}
To answer the research questions, our project collected credibility assessment labels on online medical information from both AI and human judges in Web search session contexts, and examined and compared the extent to which their judgments are cognitively biased under potential decoy effects. The data collection phase of our work consists of two components: 1) \textit{Crowdsourcing user experiment} where we observed users' behavioral patterns and credibility judgments on COVID treatment medical information under decoy and baseline conditions in Web search. 2) \textit{LLM-based evaluation experiment} that mimics the crowdsourcing user study and collects credibility judgments from multiple LLM agents under the same setting (e.g. topics, Web pages with associated queries and tasks, search result ranking/judgment sequence, decoy and baseline conditions). Data collected from these two studies allows us to assess the \textit{vulnerability} of varying LLM agents to decoy bias in information credibility assessments and also compare their degree of bias against human judgments as a baseline. The following sections will present the details of study design and data analysis. 

\subsection{Crowdsourcing User Experiment}
\subsubsection{Participants}
We utilized Prolific~\footnote{https://www.prolific.co} for our online crowdsourcing study, as it ensures access to a diverse and pre-screened participant pool while maintaining high data quality through transparent participant approval processes and fair compensation~\cite{palan2018prolific}. Five prescreeners were applied: 1) Native English speakers, 2) No professional education in medicine or related fields, 3) At least 50\% of studies completed with approval, and 4) No literacy difficulties. A total of 40 participants completed the pilot, and 617 completed the full experiment: 157 in Control A, 148 in Control B, 158 in Treatment Group A, and 154 in Treatment Group B. Of the participants, 52.7\% were female, 46.0\% male, and 1.3\% preferred not to disclose their gender. Ages ranged from 18 to 83 years (M=27.8, SD=8.5). 41.0\% had an undergraduate degree, 29.0\% a high school diploma, 17.8\% a graduate degree, and 9.7\% a community college degree. Participants primarily joined from the United Kingdom (60.6\%), the United States (25.3\%), Canada (8.4\%), and Australia (4.7\%). Our experiment was approved by the University of Oklahoma Institutional Review Board (\textbf{IRB \#}: 13527).

\subsubsection{Topics and Sessions}
To simulate the experience of evaluating pages within \textit{sessions}, we selected three \textit{COVID-19 treatment topics} that demand a relatively high level of domain knowledge, compared to common COVID topics that are widely debated in general topics: Monoclonal Antibodies, ACE and ARB inhibitors, and Hydroxychloroquine. We focused on COVID-19 treatments because 1) this is a critical medical area with significant social implications, and 2) narrowing the scope helps control for variations in behavior due to different topics, thus reducing contextual effects on decoy measures. The \textbf{\textit{questions}} corresponding to the tasks are: \textbf{1)} Can monoclonal antibodies cure COVID-19? \textbf{2)} Can ACE and ARBs worsen COVID-19? \textbf{3)} Can hydroxychloroquine treat COVID-19? 

Monoclonal antibodies are lab-created molecules designed to mimic the immune system in combating pathogens, including viruses. These antibodies were developed to target SARS-CoV-2, the virus causing COVID-19, and were used as treatment for high-risk patients. ACE inhibitors and ARBs, commonly used to treat hypertension and heart failure, gained attention during the pandemic due to their interaction with the renin-angiotensin-aldosterone system (RAAS), which also regulates blood pressure and fluid balance. Hydroxychloroquine, an antimalarial drug, was initially considered for COVID-19 treatment due to its antiviral effects seen in vitro, though later studies provided mixed and inconclusive results. We chose these specific topics from the TREC Health Misinformation track dataset~\footnote{https://trec-health-misinfo.github.io/2020.html} for two main reasons: 1) The questions related to these topics have YES/NO answers validated by external experts, which provide a reliable ground truth for assessing information credibility and distinguishing correct information from misinformation. 2) These topics involve specialized COVID-19 treatment knowledge, making them less familiar to most participants. The ground truth labels helped us reliably evaluate user judgments. We collected 40 web pages for each topic, using four predefined query-SERP combinations with ten organic search results per SERP.

\subsubsection{Decoy Design}
\begin{figure*}
\includegraphics[width=0.85\linewidth]{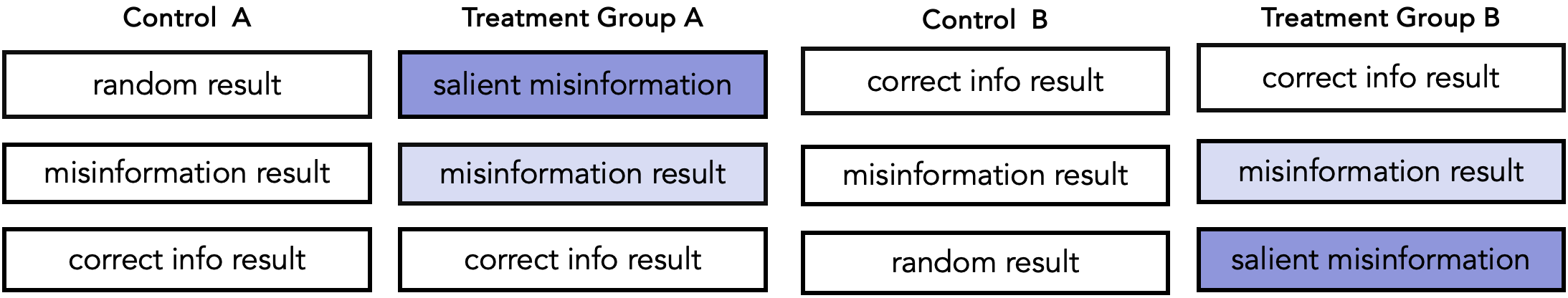}
\caption{Experimental Conditions. A decoy result was added to EACH SERP in treatment sessions. Under treatment groups, dark blue indicates decoy results, and light blue indicates target misinformation results.}
\label{fig:decoygroup}
\end{figure*}

To examine how the decoy effect can occur on SERPs, we strategically designed decoy search results by modifying the \textbf{title} and \textbf{short abstract} of snippets within the treatment groups’ SERPs. Each COVID treatment topic featured a decoy result and a target misinformation result that shared closely related subtopics relevant to the task question, while the correct information focused on a slightly different subtopic~\footnote{Although the correct information result does not share the exact subtopic with the decoy, it remains highly relevant to the task.}. The decoy result was intentionally crafted as an "obviously false" option to amplify the perceived credibility of the target misinformation. This was achieved by enriching the target's snippet with specific \textit{subject or entity} information (e.g., details about the lab or university conducting the study) and/or \textit{statistics} (e.g., the number of participants involved in the study). Research suggests that such detailed information leads users to perceive the content as higher quality and more trustworthy~\cite[e.g.][]{zhang2015quality, diviani2015low}, thereby creating a clear contrast between the target and decoy options in both user and LLM experiments.

For instance, under the topic of \textit{hydroxychloroquine}, in the treatment groups, both the decoy and target results focused on hydroxychloroquine’s effects on COVID-19 symptoms, whereas the correct result, while related to the general topic, addressed a different subtopic (e.g., hydroxychloroquine's impact on hospitalized patients). The target result was enriched with more detailed \textit{subject information} (e.g., international clinical trials, medical centers) and \textit{statistical data} (e.g., number of participants and trials, success rates, number of research teams), making it appear more credible compared to the decoy. This approach was designed to strike a balance between creating realistic conditions for studying the decoy effect and maintaining the authenticity of a natural search experience. To ensure controlled conditions, the main content of the web pages remained unchanged; the \textit{only} difference between a treatment group and its corresponding control group was the presence of the decoy result. Furthermore, to explore the influence of \textit{rank position} bias, we created two decoy effect groups: group \textbf{A}, where the decoy was ranked higher than the target misinformation, encouraging users to interact with the decoy first, and group \textbf{B}, where the target misinformation was ranked higher. Figure \ref{fig:decoygroup} outlines the four conditions, with the top three SERP positions reserved for the target, decoy, and correct results, while the remaining organic results stayed consistent across all conditions under each topic.

\subsubsection{Experimental Flow and Data Collection}
\begin{figure}
    \centering
    \includegraphics[width=0.8\linewidth]{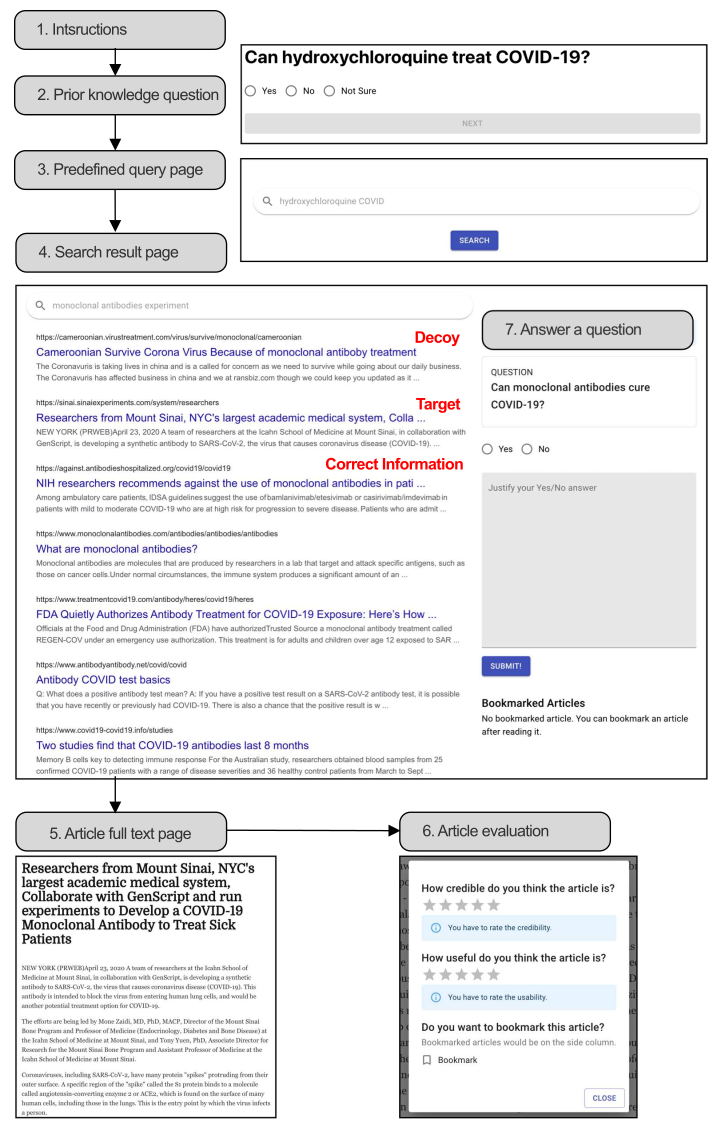}
    \caption{Crowdsourcing Experiment Flow.}
    \label{fig:flow}
\end{figure}
We collected data from 540 participants during the first phase, with 45 participants assigned to each of the 12 conditions (3 topics $\times$ 4 treatments). After excluding 60 participants who spent less than five minutes and 6 who provided identical ratings for all articles, data from 474 participants were obtained. To reach the target of 45 effective participants \textit{per condition}, we conducted a second data collection session. Participants were recruited in batches to ensure that each condition had 45 participants, ultimately resulting in 540 effective data points. The data collection process is shown in Figure~\ref{fig:flow}. 

To reasonably simulate a comparable, multi-query/SERP search session experience, we asked every participant to complete four consecutive query and SERP combinations/iterations on a given topic and click and open a minimum of three articles from the ranked results on each SERP before they could answer the related questions and finalize their session. To ensure we gathered enough data on participants' assessments of credibility, they were required to rate the credibility level of each article after their reading and before they exited the page. While participants were allowed to revise their ratings after the first read, it was not mandatory. As a result, each participant provided ratings for at least 12 articles. Participants took a median of 12.26 minutes to complete the study, with an average completion time of 14.58 minutes (SD = 7.85 minutes). Participants were compensated two US dollars each. 

\subsection{LLM-based Evaluation Experiment}
\begin{figure}
    \centering
    \includegraphics[width=0.65\linewidth]{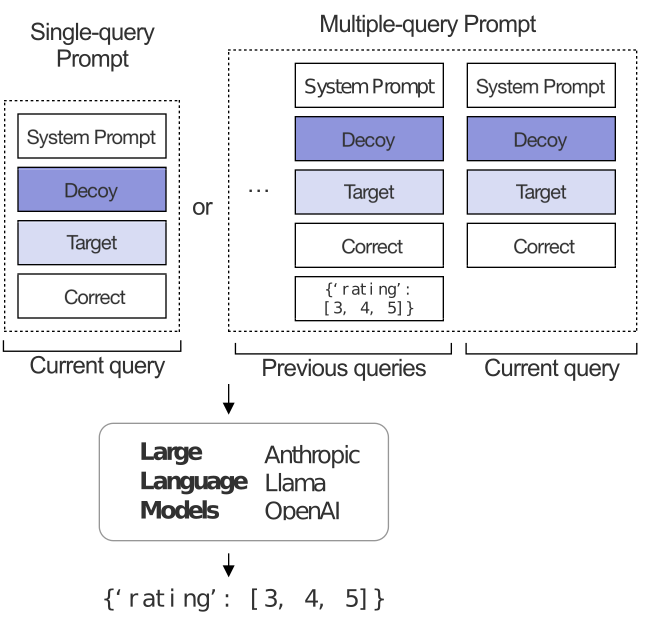}
    \caption{LLM-based Experiment Flow.}
    \label{fig:llm-flow}
\end{figure}

To assess the decoy effect and compare the credibility assessments between human and LLM judges, we replicated the crowdsourcing experiment using LLMs. The process flow of the LLM-based experiment is illustrated in Figure~\ref{fig:llm-flow}. The experiment introduces two types of prompts: single-query  and multiple-query prompts. 
\begin{enumerate}
    \item \textbf{Single-query prompts} start with a system prompt, followed by three articles retrieved by the current query: a decoy article in the treatment group or a random article in the control group, a target article, and a correct article.
    \item \textbf{Multiple-query prompts} start by providing a series of previous queries (each task includes four queries) and their corresponding credibility ratings produced by the LLM for those queries. This simulates a multi-query, session-wide context, where the current query is appended to the end of the sequence. This context gives the LLM access to the entire sequence of prior interactions within the same search session. The content provided for each current query is the same as that in the single-query prompts.
\end{enumerate}
The main difference between the multiple-query prompts and the single-query prompts is that for the subsequent queries following the first query in the multiple-query prompts, the LLM can take its own past judgment into account, which may influence its current judgment. This allows us to examine how credibility assessments and cognitive biases of LLMs evolve across a series of related queries. Also, the multi-query and session-wide context simulated by the multiple-query prompts are similar to the crowdsourcing user experiment.

We evaluated LLMs' credibility judgments in both prompts for two key reasons: 1) Previous research on document judgments mainly focus on ad hoc retrieval or single-query evaluation contexts (e.g.~\citep{eickhoff2018cognitive, scholer2013effect, sakai2020retrieval}), leaving judgments under whole-session/multi-query contexts generally understudied~\citep{liu2022toward}; 2) LLMs tend to respond differently when making multiple judgments within the same conversation versus separate, individual ones. Considering both types of prompts allow us to obtain a deeper understanding of LLMs' ``cognitive biases'' under varying evaluation settings. 

The system prompt structure includes two parts: 

\textbf{[Role]} \textit{As an information searcher and non-medical expert, you are trying to answer the question about COVID-19: '\{task\}' by searching for and using relevant information retrieved online.} 

\textbf{[Task]} \textit{You will read related documents retrieved under \{nth\} query "\{query\}" in the search session under this task and evaluate their credibility. You will be provided with the titles and full text of three articles. After reading the three articles, please rate each article's credibility using a 5-point scale. Please note that the topical relevance of the article to the given query or the task question is different from the credibility of the article. Rate solely on the credibility of the article, disregarding its topical relevance to the given query or task question. Return the ratings in JSON format as an array of three numbers under the key of "rating". No explanation.}

\{task\}, \{nth\}, and \{query\} are variables that are substituted based on the topic and query.

For multiple-query or whole-session prompts, the [Role] part would be used only once at the beginning of each LLM session (four queries), while the [Task] part would be used for each query. For a single-query prompt, both [Role] and [Task] would be used at each session (one query). However, in single-query prompts, the first sentence of the [Task] section becomes: \textit{You will evaluate the credibility of articles retrieved under the query "{query}" in this search session}, since no prior queries are referenced.

In our study, we utilized a diverse selection of nine LLMs to evaluate decoy effects in information judgments across multiple tasks or subtopics of COVID treatment. These models were chosen not only for their architecture and parameter sizes but also because they span a range of generation capabilities and training data. The diverse selection of models ensures that our study captures a broad spectrum of performance and responses to potential biases. To maintain consistency and comparability with the crowdsourcing user experiments, each model runs each prompt 45 times. The LLM models included in the experiment are as follows:
\begin{itemize}
\item Llama-3.1-70B: An open-source AI model was released in July 2024. The training data are up to December 2023. Due to limited computing resources, we used the EXL2 quantization version of the Llama-3.1-70B Instruct \footnote{https://huggingface.co/turboderp/Llama-3.1-70B-Instruct-exl2/tree/4.5bpw}.
\item Llama-3-70B: An earlier version of the Llama series, released in April 2024. The training data are up to December 2023. We also used a quantized version \footnote{https://huggingface.co/turboderp/Llama-3-70B-Instruct-exl2/tree/4.5bpw}.
\item Llama-3-8B: A smaller model in the Llama series with 8 billion parameters. The training data have a cutoff date of March 2023 \footnote{https://huggingface.co/turboderp/Llama-3-8B-Instruct-exl2/tree/5.0bpw}.
\item gpt-4o: OpenAI's current flagship model. The snapshot used in this study is gpt-4o-2024-05-13. The training data are up to October 2023.
\item gpt-4o-mini: A small and cheaper model for light tasks offered by OpenAI. The snapshot used in this study is gpt-4o-mini-2024-07-18. The training data are up to October 2023.
\item gpt-3.5-turbo: The snapshot used in the study is gpt-3.5-turbo-0125. The training data are up to September 2021.
\item claude-3.5-sonnet: The latest model in the Claude series. The training data cut-off date is April 2024. The API model we used is claude-3-5-sonnet-20240620.
\item claude-3-sonnet: Another model in the Claude series. The training data cut-off date is August 2023. The API model we used is claude-3-sonnet-20240229.
\item claude-3-haiku: The fastest but most compact model in the Claude series. The training data cut-off date is August 2023. The API model we used is claude-3-haiku-20240307.
    
\end{itemize}

We used OpenAI and Anthropic APIs for accessing GPT models and Claude models, respectively. The Llama models were deployed locally on a system running Ubuntu 22.04.4 LTS, equipped with an Intel i9-10920X CPU (24 threads), two NVIDIA Quadro RTX 6000 GPUs with 24GB of memory each, and 128 GB of RAM. Exllamav2\footnote{https://github.com/turboderp/exllamav2}, a Python fast inference library for running LLMs, was used to run the Llama models.

\section{Results}
This section presents the results from our user and LLM experiments in response to the proposed RQs on human and AI assessments. 

\subsection{RQ1: Decoy effect on the Credibility Assessments of LLM and Human Judges}

Since all the crowdsourcing participants complete four queries in their studies, we only compared results from LLM-based experiments using multiple-query prompts with human judges here.

\subsubsection{Credibility Assessment}

\begin{figure*}
\includegraphics[width=1.05\linewidth]{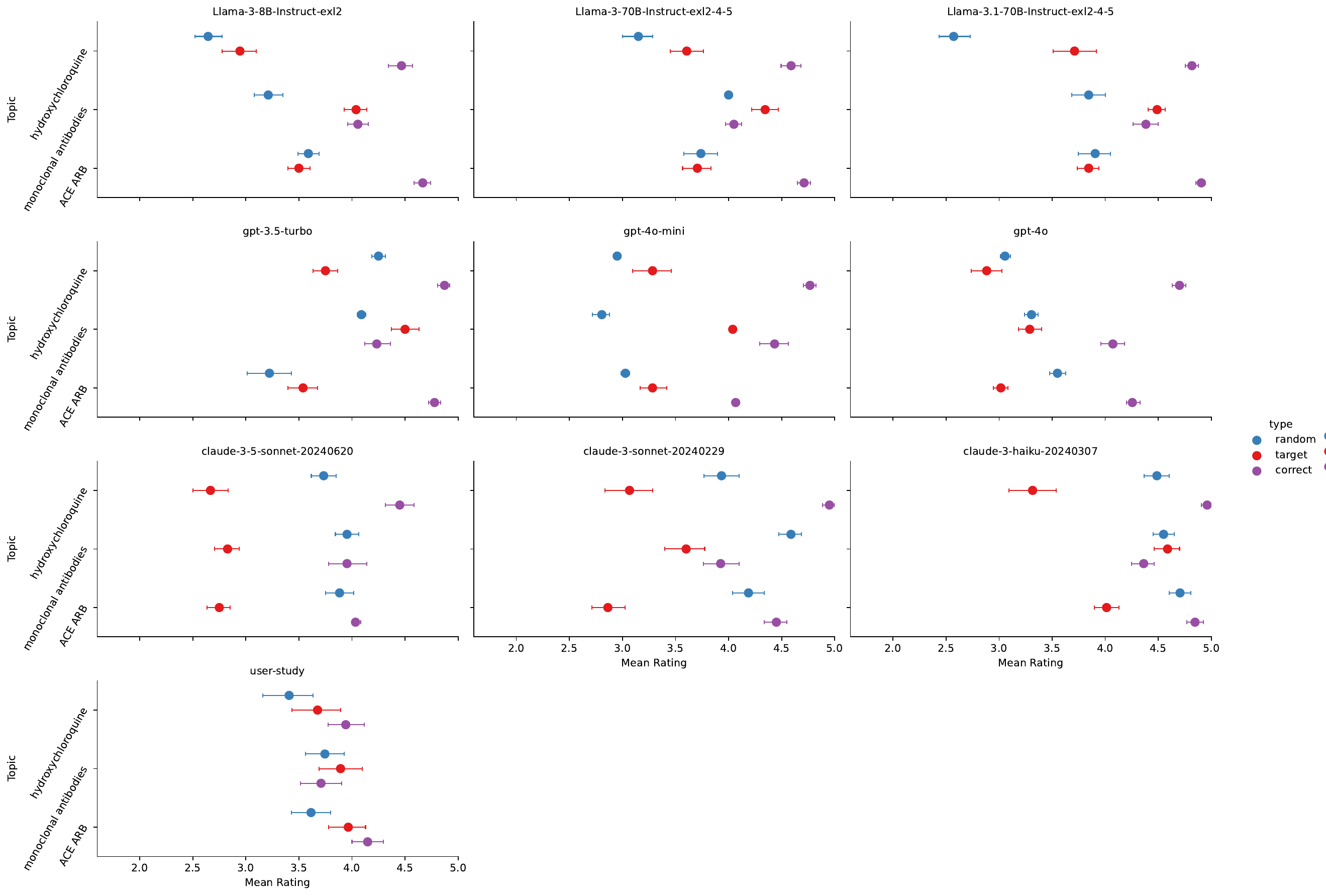}
\caption{Mean ratings by topics and article types for the control groups.}
\label{fig:mean_rating_by_topic_and_article_type_control}
\end{figure*}

The results in Figure~\ref{fig:mean_rating_by_topic_and_article_type_control} present the mean ratings by topic and article type for the control group across various models. Each subplot represents a different model, and the ratings are categorized by three types: random (blue), target (red), and correct (purple). The bars represent 95\% confidence intervals.

For crowdsourcing participant ratings from the user study, the correct responses received the highest mean ratings for two of the three topics (hydroxychloroquine and ACE ARB), followed by target and random responses. This pattern suggests that the participants were generally able to differentiate between correct information and misleading information, though some inconsistencies were observed. More recently released GPT and Claude models, including  GPT-4O, GPT-4O-mini, Claude-3.5-Sonnet, and Claude-3-Sonnet, demonstrated a significantly more consistent and effective ability to distinguish correct information from both random responses and target misinformation. In these models, there was a clear and noticeable gap in ratings between correct information and other types of content. However, other models also showed a clear distinction between correct information and target misinformation in general but displayed weakness in assessing correct information about ``monoclonal antibodies''. The models assigned low scores to correct information about ``monoclonal antibodies'' compared to target misinformation. This discrepancy could be related to the models' knowledge cut-off dates or inherent performance limitations of earlier large language models.

Regarding the ratings of random information in control groups, results varies greatly across models. Some models, such as the Llama series, GPT-3.5, and GPT-4O-mini, consistently assign lower ratings to random information compared to other types of information. This is likely because random articles often contain credible information, but with limited relevance to the specific query or topic. These models may struggle to clearly distinguish between relevance and credibility, even when provided with explicit guidance to assess both. In contrast, models from the Claude series, particularly Claude-3.5-Sonnet and Claude-3-Sonnet, exhibit a more nuanced ability to differentiate between types of information. These models demonstrate a clear separation between misinformation (target) and credible information (whether random or correct), consistently evaluating random content accurately based on its credibility.

\begin{figure*}
\includegraphics[width=1.05\linewidth]{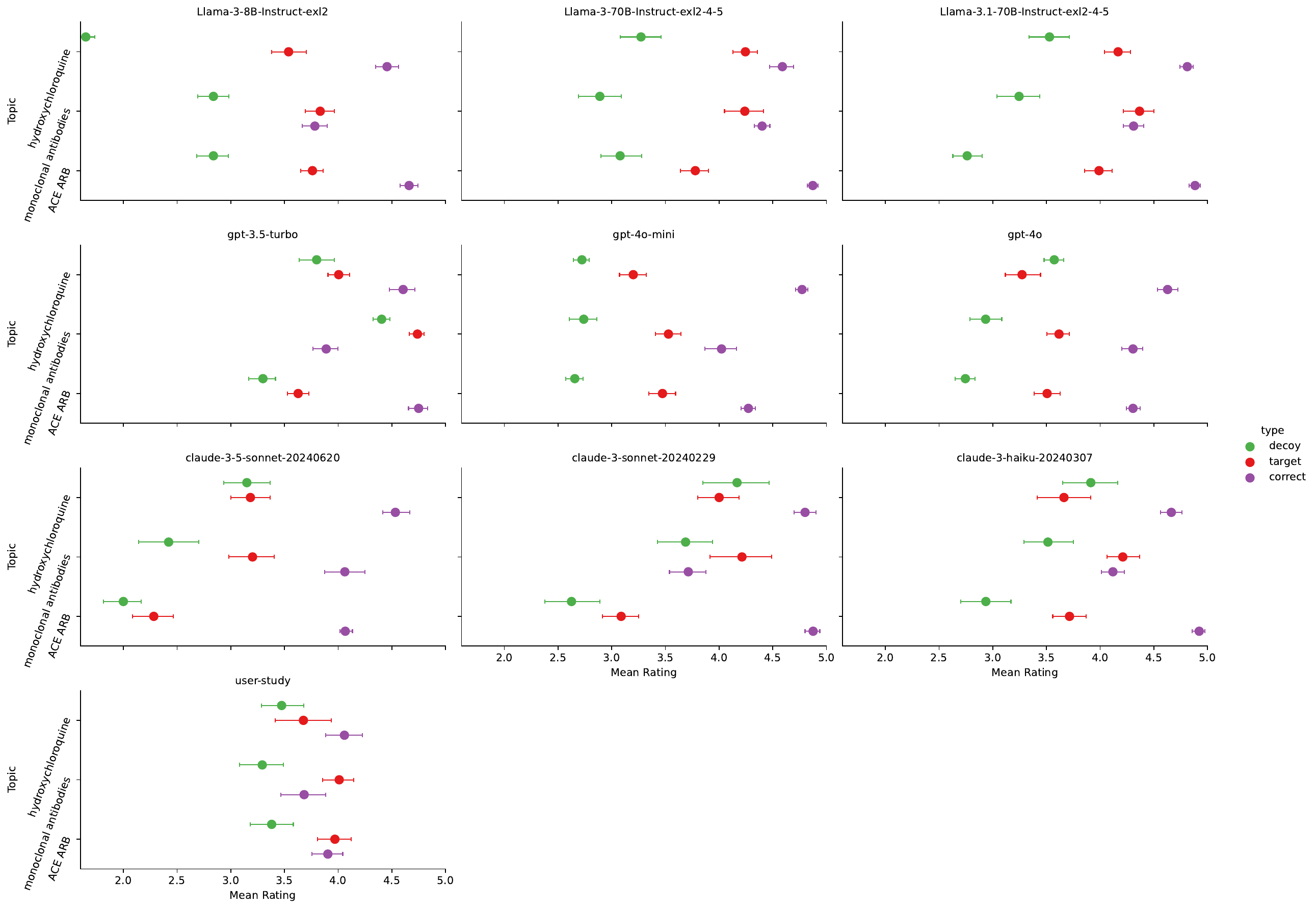}
\caption{Mean ratings by topics and article types for the treatment groups.}
\label{fig:mean_rating_by_topic_and_article_type_treatment}
\end{figure*}

Figure \ref{fig:mean_rating_by_topic_and_article_type_treatment} presents the mean ratings for different topics and article types in the treatment groups across various models. Each subplot represents a different model, with ratings categorized under three types: decoy (green), target (red), and correct (purple). For the crowdsourcing user study, decoy articles consistently received lower ratings compared to target and correct information, which aligns with the intended design of the decoy articles, presenting salient misinformation. However, concerning correct articles, participant ratings were not always consistently higher than those for decoy and target articles presenting misinformation. 

Most of the LLMs performed well in distinguishing decoy articles from correct and target ones, with decoy responses generally receiving the lowest ratings. Llama and GPT models demonstrated a clearer separation of decoy articles compared to Claude models. When it comes to correct responses, all models consistently rated them the highest on average, showing that the models were generally effective at identifying accurate information. However, the degree of separation between correct and target ratings varied slightly across different topics, particularly for the topic ``monoclonal antibodies.'' Only recent models, such as GPT-4O-mini, GPT-4O, and Claude-3.5-Sonnet, performed consistently across all topics, maintaining a clearer separation between correct and target responses.

The confidence intervals in Figures \ref{fig:mean_rating_by_topic_and_article_type_control} and \ref{fig:mean_rating_by_topic_and_article_type_treatment}, represented by the bars, suggest that crowdsourcing participants exhibited more variability in their ratings compared to the LLMs. This is expected, as human raters might interpret and score content more diversely, leading to a broader range of ratings. Overall, more recent GPT and Claude models showed greater consistency and accuracy in distinguishing credible information (correct and random) from misinformation (decoy and target). Among all the models, the credibility assessment of GPT-4o, Claude-3.5-Sonnet, and Claude-3-Sonnet aligned most closely with the objectives of our study design and article selection criteria. Specifically, in our design, correct information is characterized by high credibility, while random information is also credible, but with lower relevance to the query. On the other hand, target information contains misinformation, and decoy articles include misinformation with more saliently misleading elements than target articles. These models effectively captured these distinctions, consistently rating correct information the highest credibility, random information as credible, and identifying misinformation in both the decoy and target information with lower ratings. 

\subsubsection{Compare Decoy Effects between LLMs and Human Judges}\label{sec:compare-decoy}

Next, we compare the decoy effects between the results from the crowdsourcing experiment and the LLM-based experiment. Decoy effects can be assessed by examining the differences in ratings assigned to target articles with misinformation between the control group (where no decoy article was present) and the treatment group (where a decoy article with salient misinformation was presented alongside the target article). A t-test was conducted to determine if the observed differences in ratings between the two groups are statistically significant.

\begin{table}[h!]
\centering
\resizebox{0.7\columnwidth}{!}{%
\begin{tabular}{l l l l l l l}
\toprule
\multirow{2}{*}{model} & \multirow{2}{*}{topic} & \multicolumn{2}{c}{Control} & \multicolumn{2}{c}{Treatment} & \multirow{2}{*}{t-test} \\
\cmidrule(lr){3-4} \cmidrule(lr){5-6}
 &  & random & target & decoy & target &  \\
\midrule
\multirow{3}{*}{Llama-3.1-70B} & 1 & 3.91 (1.04) & 3.84 (0.72) & 2.76 (0.96) & 3.99 (0.85) & -1.74 \\
 & 2 & 2.57 (1.05) & 3.71 (1.39) & 3.53 (1.20) & 4.17 (0.80) & \textbf{-3.80}*** \\
 & 3 & 3.84 (1.17) & 4.49 (0.56) & 3.24 (1.42) & 4.37 (0.94) & 1.49 \\
\midrule
\multirow{3}{*}{Llama-3-70B} & 1 & 3.74 (1.04) & 3.71 (0.92) & 3.08 (1.24) & 3.78 (0.90) & -0.75 \\
 & 2 & 3.15 (0.98) & 3.61 (1.05) & 3.27 (1.27) & 4.24 (0.80) & \textbf{-6.49}*** \\
 & 3 & 4.00 (0.00) & 4.34 (0.83) & 2.89 (1.43) & 4.24 (1.26) & 0.94 \\
\midrule
\multirow{3}{*}{Llama-3-8B} & 1 & 3.59 (0.68) & 3.50 (0.74) & 2.84 (0.98) & 3.76 (0.76) & \textbf{-3.28}** \\
 & 2 & 2.64 (0.88) & 2.94 (1.14) & 1.65 (0.57) & 3.54 (1.10) & \textbf{-5.05}*** \\
 & 3 & 3.21 (0.90) & 4.04 (0.73) & 2.84 (1.00) & 3.83 (0.98) & 2.25* \\
\midrule
\multirow{3}{*}{gpt-4o} & 1 & 3.55 (0.50) & 3.02 (0.48) & 2.74 (0.62) & 3.51 (0.83) & \textbf{-6.85}*** \\
 & 2 & 3.06 (0.35) & 2.88 (0.99) & 3.57 (0.64) & 3.27 (1.05) & \textbf{-3.61}*** \\
 & 3 & 3.31 (0.46) & 3.29 (0.79) & 2.93 (1.01) & 3.62 (0.71) & \textbf{-4.12}*** \\
\midrule
\multirow{3}{*}{gpt-4o-mini} & 1 & 3.03 (0.29) & 3.28 (0.83) & 2.66 (0.55) & 3.47 (0.87) & \textbf{-2.11}* \\
 & 2 & 2.95 (0.22) & 3.28 (1.30) & 2.72 (0.51) & 3.20 (0.90) & 0.71 \\
 & 3 & 2.81 (0.52) & 4.04 (0.19) & 2.74 (0.95) & 3.53 (0.84) & 8.00*** \\
\midrule
\multirow{3}{*}{gpt-3.5-turbo} & 1 & 3.22 (1.49) & 3.54 (1.01) & 3.30 (0.86) & 3.63 (0.65) & -1.00 \\
 & 2 & 4.25 (0.43) & 3.75 (0.83) & 3.80 (1.14) & 4.01 (0.71) & \textbf{-3.13}** \\
 & 3 & 4.09 (0.29) & 4.50 (0.86) & 4.41 (0.53) & 4.74 (0.45) & \textbf{-3.29}** \\
\midrule
\multirow{3}{*}{claude3-5-sonnet} & 1 & 3.88 (0.56) & 2.75 (0.44) & 2.00 (0.71) & 2.28 (0.83) & 3.87*** \\
 & 2 & 3.73 (0.48) & 2.67 (0.66) & 3.15 (0.84) & 3.18 (0.75) & \textbf{-4.03}*** \\
 & 3 & 3.95 (0.45) & 2.83 (0.49) & 2.42 (1.17) & 3.20 (0.89) & \textbf{-2.94}** \\
\midrule
\multirow{3}{*}{claude3-sonnet} & 1 & 3.16 (0.95) & 4.06 (0.76) & 2.73 (1.11) & 4.02 (0.95) & \textbf{-1.98}* \\
 & 2 & 2.59 (1.10) & 3.83 (1.16) & 3.45 (0.96) & 4.39 (0.75) & \textbf{-5.86}*** \\
 & 3 & 3.13 (0.80) & 4.41 (0.71) & 2.68 (1.40) & 4.07 (1.17) & \textbf{-3.51}*** \\
\midrule
\multirow{3}{*}{claude3-haiku} & 1 & 4.71 (0.46) & 3.71 (0.92) & 3.08 (1.24) & 3.78 (0.90) & 2.90** \\
 & 2 & 3.15 (0.98) & 3.61 (1.05) & 3.27 (1.27) & 4.24 (0.80) & -1.78 \\
 & 3 & 4.00 (0.00) & 4.34 (0.83) & 2.89 (1.43) & 4.24 (1.26) & 3.83*** \\

\bottomrule
\end{tabular}
}

\caption{Decoy effects in multiple-query sessions with mean ratings, standard deviations, and t-test statistics. Topics: 1 = ACE ARB, 2 = hydroxychloroquine, 3 = monoclonal antibodies. Significance levels: * p<0.05, ** p<0.01, *** p<0.001.}
\label{table:multiple_query_decoy}
\end{table}

Table~\ref{table:multiple_query_decoy} presents the decoy effects observed during multiple-query sessions, showing the mean ratings, standard deviations, and t-test statistics across various models and topics. The decoy effect can be detected by comparing the target article ratings in the treatment group against the control group. There is a decoy effect occurred only when the target article's rating in the treatment group is significantly higher than in the control group, i.e., an increase in target rating due to the presence of the decoy. We observe that target ratings in the treatment group are often significantly higher than in the control group, suggesting that the presence of decoy articles positively influences the perceived credibility of the target articles. This implies that decoy articles, which contain more salient misinformation than target ones, can manipulate LLMs' assessment, which makes misinformation in target articles appear more credible than it would be with the decoy.

Interestingly, the models that are more likely affected by decoy effects tend to be the more recent and larger LLMs, such as GPT-4O and Claude-3-Sonnet, which are also the models that generally performed better in credibility assessments, as illustrated in Figures~\ref{fig:mean_rating_by_topic_and_article_type_control} and \ref{fig:mean_rating_by_topic_and_article_type_treatment}. This suggests that while these models are better at identifying correct information, they may also be more vulnerable to cognitive biases triggered by the presence of strongly misleading decoy articles. On the other hand, Llama models appear less affected by the presence of decoy articles, exhibiting smaller differences in target ratings between control and treatment groups. This suggests that while Llama models might be less prone to decoy-driven shifts in assessment. These findings suggests that models that are better at distinguishing between credible and incredible information might also be more vulnerable to subtle manipulative tactics, like the presence of a decoy article.

\begin{table}[h!]
\footnotesize
\centering
\resizebox{0.6\columnwidth}{!}{%
\begin{tabular}{l l l l l l}
\toprule
\multirow{2}{*}{Topic} & \multicolumn{2}{c}{Control} & \multicolumn{2}{c}{Treatment} & \multirow{2}{*}{t-test} \\
\cmidrule(lr){2-3} \cmidrule(lr){4-5}
 & random & target & decoy & target &  \\
\midrule
\multicolumn{6}{c}{All query sessions} \\
\midrule
1 & 3.57 (1.03) & 3.95 (0.88) & 3.38 (1.18) & 3.98 (0.79) & -0.37 \\
2 & 3.66 (0.93) & 3.68 (0.98) & 3.40 (1.22) & 3.62 (1.08) & 0.49 \\
3 & 3.71 (0.92) & 3.96 (0.99) & 3.40 (1.19) & 4.01 (0.78) & -0.53 \\
\midrule
\multicolumn{6}{c}{Clicked on decoy/random prior to target} \\
\midrule
1 & 3.57 (1.03) & 3.95 (0.88) & 2.81 (1.53) & 4.44 (0.80) & \textbf{-2.95}** \\
2 & 3.66 (0.93) & 3.68 (0.98) & 3.36 (1.18) & 3.82 (0.85) & -0.63 \\
3 & 3.71 (0.92) & 3.96 (0.99) & 3.73 (1.16) & 3.82 (1.22) & 0.65 \\

\midrule
\multicolumn{6}{c}{Clicked on all the first three articles} \\
\midrule
1 & 3.49 (0.98) & 4.02 (0.92) & 3.38 (1.14) & 3.83 (0.80) & 1.59 \\
2 & 3.63 (1.01) & 3.56 (1.01) & 3.30 (1.25) & 3.56 (1.08) & 0.03 \\
3 & 3.69 (1.16) & 3.62 (1.24) & 3.33 (1.21) & 4.03 (0.68) & \textbf{-2.06}* \\
\midrule
\multicolumn{6}{c}{Participants without prior knowledge} \\
\midrule
1 & 3.69 (1.11) & 3.91 (0.89) & 3.40 (1.12) & 3.98 (0.76) & -0.67 \\
2 & 3.73 (0.88) & 3.97 (0.69) & 3.50 (1.21) & 3.40 (1.17) & 3.67*** \\
3 & 3.81 (0.73) & 3.97 (0.90) & 3.29 (1.15) & 3.89 (0.81) & 0.79 \\
\midrule
\multicolumn{6}{c}{Participants with prior knowledge} \\
\midrule
1 & 3.35 (0.82) & 4.03 (0.84) & 3.32 (1.36) & 4.00 (0.89) & 0.18 \\
2 & 3.61 (0.96) & 3.43 (1.12) & 3.24 (1.23) & 3.97 (0.82) & \textbf{-3.12}** \\
3 & 3.58 (1.13) & 3.95 (1.12) & 3.67 (1.26) & 4.22 (0.69) & -1.85 \\
\bottomrule
\end{tabular}
}
\caption{Decoy effects in crowdsourcing experiment with mean ratings, standard deviations, and t-test statistics. Topics: 1 = ACE ARB, 2 = hydroxychloroquine, 3 = monoclonal antibodies. Significance levels: * p<0.05, ** p<0.01, *** p<0.001.}
\label{table:decoy_crowdsourcing}
\end{table}


We also analyzed the data collected from the crowdsourcing study, which is presented the data in Table~\ref{table:decoy_crowdsourcing}. This table shows the mean ratings, standard deviations, and t-test statistics for decoy effects across various search contexts. The results suggest that decoy effects are not consistently observed when considering all query sessions. This may be because not all participants clicked on the decoy article, even though decoy articles were the first presented in the treatment group, thus limiting the opportunity for the decoy effect to influence their ratings.

To investigate decoy effects more closely, we examined query iterations where participants either clicked on a random article in the control group or clicked on a decoy article in the treatment group (\textbf{Clicked on decoy/random prior to target}). A clear decoy effect was observed for Topic 1 (ACE ARB), where the target article's rating in the treatment group (4.44) was significantly higher than the rating in the control group (3.95) with a t-test value of -2.95 (p < 0.01). This suggests that when participants clicked on the decoy article first, their subsequent perception of the target article was significantly influenced. However, this effect was not observed for Topics 2 (hydroxychloroquine) or 3 (monoclonal antibodies).

In the context of \textbf{Clicked on All the First Three Articles}, we included query data where participants clicked on the decoy, target, and correct articles in the treatment group, or clicked on random, target, and correct articles in the control group. A significant decoy effect was seen for Topic 3 (monoclonal antibodies), where the target article rating in the treatment group (4.03) was significantly higher than in the control group (3.62), with a t-test value of -2.06 (p < 0.05). However, no significant decoy effects were found for Topics 1 and 2 in this context.

Prior knowledge may also affect users' perception of credibility. We separate data into two groups based on whether participants have prior knowledge on the topic based on a pre-study question. For participants without prior knowledge, no decoy effect is observed. For participants with prior knowledge, a significant decoy effect is observed for Topic 2 (hydroxychloroquine). The target rating in the treatment group (3.97) is significantly higher than in the control group (3.43), with a t-test value of -3.12 (p < 0.01). But Topic 2 is the only topic where we observed decoy effect.

We also explored whether prior knowledge of the topic affected participants' perceptions of credibility.Participants were grouped based on their responses to a pre-study question (see Figure \ref{fig:flow}) that asked about prior knowledge. If they responded ``Yes'' or ``No,'' they were classified as having prior knowledge of the topic. If they answered ``Not sure,'' they were classified as not having prior knowledge. For participants without prior knowledge, no significant decoy effects were observed for any of the three topics. This suggests that participants unfamiliar with the topic were less influenced by the presence of decoy articles. However, for participants with prior knowledge, we observed a significant decoy effect for Topic 2 (hydroxychloroquine). In this case, the target rating in the treatment group (3.97) was significantly higher than the control group rating (3.43), with a t-test value of -3.12 (p < 0.01). This indicates that when the decoy articles were present, participants with prior knowledge tended to increase their ratings of the target article.

In summary, the strength of the decoy effects varies between different topics, participant behaviors, and search contexts. Significant decoy effects were observed in specific scenarios, such as when participants clicked on the decoy article first, or when participants had prior knowledge of the topic. Compared to the LLM-based experiment, where LLM models demonstrated more consistent rating patterns and were generally more prone to the influence of decoy articles, decoy effects on human participants appear to be more contextually dependent. Factors such as participants' behavior during search tasks (e.g., whether they clicked on the decoy article) and their prior knowledge of the topic play a key role in shaping how they perceive the credibility of information. Thus, LLMs may demonstrate vulnerability to decoy effects more broadly compared to human participants.

\subsection{RQ2: Decoy Effect on LLM Agents under Varying Assessment Contexts}
In Section~\ref{sec:compare-decoy}, we analyzed the results of the LLM-based experiment using multiple-query prompts (see Figure \ref{fig:llm-flow}), where multiple queries were presented in a single prompt, maintaining context across the queries (there are four queries for each topic). However, chat-based large language models (LLMs) can also process information without retaining conversational context between queries, which can affect their performance. Thus, we conducted another LLM-based experiment by using single-query prompts, which means articles in a query would be in separate query from other three queries for a topic and each query would be a single prompt to LLMs.

\begin{table}[h!]
\centering
\resizebox{0.7\columnwidth}{!}{%

\begin{tabular}{l l l l l l l}
\toprule
\multirow{2}{*}{Model} & \multirow{2}{*}{Topic} & \multicolumn{2}{c}{Control} & \multicolumn{2}{c}{Treatment} & \multirow{2}{*}{t-test} \\
\cmidrule(lr){3-4} \cmidrule(lr){5-6}
 &  & random & target & decoy & target &  \\
\midrule
\multirow{3}{*}{Llama-3.1-70B} & 1 & 2.73 (0.94) & 4.22 (0.56) & 2.79 (1.05) & 4.04 (0.82) & 2.46* \\
 & 2 & 2.21 (0.96) & 3.86 (1.41) & 3.61 (1.12) & 4.38 (0.72) & \textbf{-4.36}*** \\
 & 3 & 3.33 (0.93) & 4.31 (0.65) & 2.99 (1.57) & 4.43 (0.85) & -1.45 \\
\midrule
\multirow{3}{*}{Llama-3-70B} & 1 & 3.16 (0.95) & 4.06 (0.76) & 2.73 (1.11) & 4.02 (0.95) & 0.49 \\
 & 2 & 2.59 (1.10) & 3.83 (1.16) & 3.45 (0.96) & 4.39 (0.75) & \textbf{-5.51}*** \\
 & 3 & 3.13 (0.80) & 4.41 (0.71) & 2.68 (1.40) & 4.07 (1.17) & 3.39*** \\
\midrule
\multirow{3}{*}{Llama-3-8B} & 1 & 3.23 (0.91) & 3.78 (0.86) & 2.48 (1.05) & 3.93 (0.68) & -1.84 \\
 & 2 & 2.55 (1.07) & 3.73 (1.31) & 3.27 (1.19) & 4.27 (0.69) & \textbf{-4.85}*** \\
 & 3 & 2.68 (0.82) & 4.23 (0.67) & 2.14 (0.88) & 3.85 (1.04) & 4.16*** \\
\midrule
\multirow{3}{*}{gpt-4o} & 1 & 2.97 (0.28) & 3.06 (0.97) & 3.03 (0.60) & 3.14 (0.92) & -0.84 \\
 & 2 & 3.07 (0.44) & 3.29 (0.93) & 3.26 (0.62) & 3.44 (0.95) & -1.57 \\
 & 3 & 3.04 (0.22) & 3.26 (0.85) & 2.87 (0.70) & 3.54 (1.16) & \textbf{-2.70}** \\
\midrule
\multirow{3}{*}{gpt-4o-mini} & 1 & 2.55 (0.50) & 3.51 (0.86) & 2.68 (0.85) & 3.50 (1.13) & 0.05 \\
 & 2 & 2.43 (0.55) & 3.75 (1.09) & 3.07 (0.82) & 3.68 (1.11) & 0.58 \\
 & 3 & 2.64 (0.48) & 3.55 (0.86) & 2.46 (1.15) & 3.47 (1.48) & 0.61 \\
\midrule
\multirow{3}{*}{gpt-3.5-turbo} & 1 & 3.53 (0.83) & 4.19 (0.84) & 1.96 (0.68) & 3.77 (0.42) & 5.93*** \\
 & 2 & 4.00 (0.00) & 4.50 (0.87) & 3.41 (1.22) & 3.81 (0.84) & 7.63*** \\
 & 3 & 4.25 (0.43) & 5.00 (0.00) & 3.47 (1.47) & 4.51 (0.50) & 13.08*** \\
\midrule
\multirow{3}{*}{claude3-5-sonnet} & 1 & 3.83 (0.46) & 2.39 (0.49) & 2.00 (0.71) & 2.50 (1.13) & -0.71 \\
 & 2 & 3.52 (0.85) & 3.13 (0.93) & 3.00 (0.71) & 3.00 (0.71) & 0.85 \\
 & 3 & 3.56 (0.50) & 2.58 (0.56) & 2.25 (0.84) & 3.07 (0.97) & \textbf{-5.88}*** \\
\midrule
\multirow{3}{*}{claude3-sonnet} & 1 & 3.63 (0.49) & 2.28 (0.61) & 3.00 (0.93) & 2.94 (1.07) & \textbf{-4.05}*** \\
 & 2 & 2.95 (0.57) & 3.55 (1.06) & 3.33 (0.82) & 3.65 (0.94) & -0.55 \\
 & 3 & 3.65 (0.63) & 4.02 (1.10) & 3.60 (1.08) & 3.63 (1.19) & 1.82 \\
\midrule
\multirow{3}{*}{claude3-haiku} & 1 & 3.97 (0.65) & 3.78 (0.68) & 2.83 (1.23) & 3.50 (0.57) & 2.38* \\
 & 2 & 3.81 (0.48) & 3.40 (1.07) & 3.50 (0.95) & 3.68 (0.87) & -1.56 \\
 & 3 & 3.98 (0.23) & 4.34 (0.54) & 3.53 (1.06) & 4.10 (0.52) & 2.43* \\

\bottomrule
\end{tabular}
}
\caption{Decoy effects in single-query iterations with mean ratings, standard deviations, and t-test statistics. Topics: 1 = ACE ARB, 2 = hydroxychloroquine, 3 = monoclonal antibodies. Significance levels: * p<0.05, ** p<0.01, *** p<0.001.}
\label{table:decoy-single-query}
\end{table}

As shown in Table~\ref{table:decoy-single-query}, the decoy effects were far less prevalent in this single-query setup compared to the multiple-query experiment (see Table~\ref{table:multiple_query_decoy}). No model showed decoy effects for more than one topic in the single-query format. Decoy effects were notably reduced, suggesting that LLMs are less susceptible to decoy manipulation when context is not carried over between queries. While some decoy effects were still observed in isolated cases for certain models (e.g., Llama-3.1-70B and Claude3-haiku for Topic 1, and Llama models for Topic 2), overall, the single-query structure appeared to mitigate this vulnerability. This indicates that LLMs' ability to maintain context plays a substantial role in their susceptibility to decoy articles. Reducing contextual carryover between queries could be a useful strategy to minimize the impact of decoy effects.

\begin{figure}
    \centering
    \includegraphics[width=0.8\linewidth]{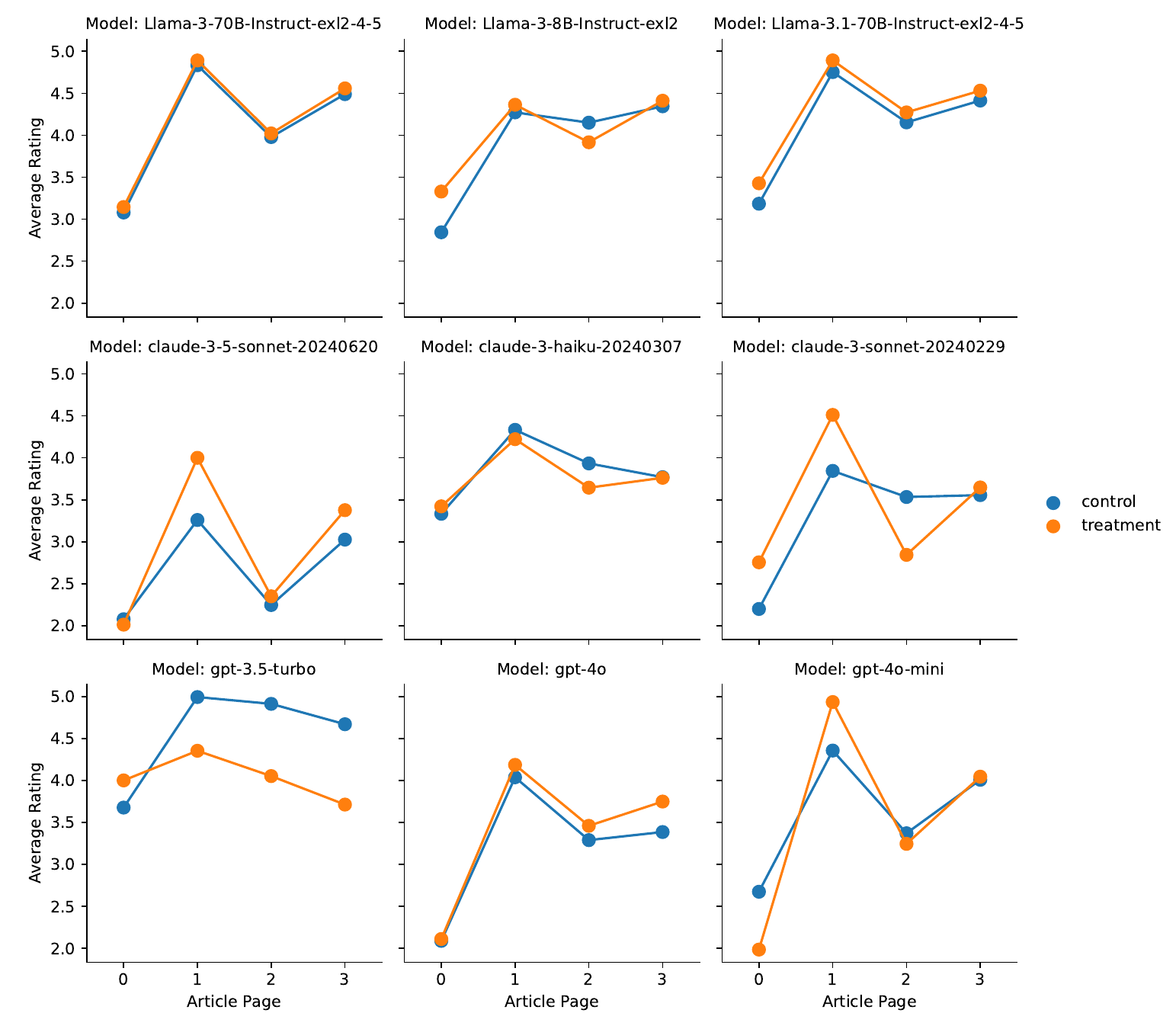}
    \caption{Average ratings for target articles by query/SERP sequence in single-query iterations, where each query-SERP combination is assessed seperately without session memory. Page 0 indicates the SERP retrieved under the first query of a search session.}
    \label{fig:page-rating-single}
\end{figure}

\begin{figure}
    \centering
    \includegraphics[width=0.8\linewidth]{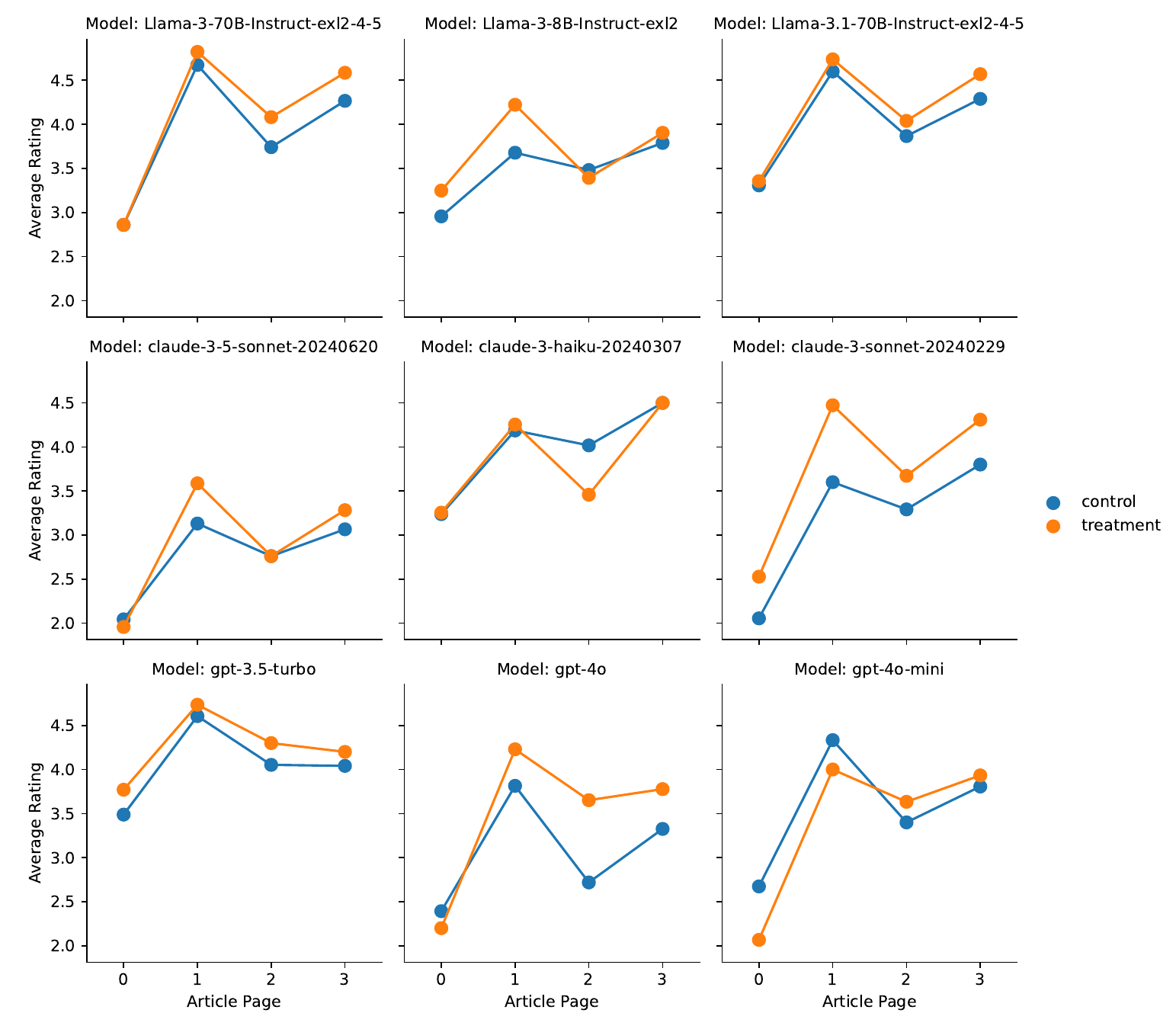}
    \caption{Average ratings for target articles by query/SERP sequence in multiple-query iterations. }
    \label{fig:page-rating}
\end{figure}

To better understand how session contexts, with memory of previous queries, enhanced decoy effects, we analyzed the ratings for target articles by query/SERP sequence in both the single-query and multiple-query experiments. The comparison is shown in Figures~\ref{fig:page-rating-single} and \ref{fig:page-rating}. Each subplot in these figures represents a different model. By comparing the two figures, we observe that the ratings for target articles on Page 0 (i.e., the first SERP) are very similar between the single-query and multiple-query setups, since there is no interaction context or session memory in either scenario at this point. However, the credibility ratings on target results in the treatment group vary greatly between single-query and multiple-query prompts, while the ratings for target articles in the control group remain similar between using the two types of prompts. There are only two models present similar trends over pages between the two types of prompts: Llama-3.1-70B and Claude-3-5-sonnet. GPT-4o, Claude-3-sonnet, and GPT-3.5-turbo present the most dramatic difference between the single-query and whole-session (multiple-query) prompts, which explained their different decoy effects presented in Table~\ref{table:multiple_query_decoy} (for multiple-query prompts) and Table~\ref{table:decoy-single-query} (for single-query prompts). These results suggest that session memory and interaction context in multiple-query prompts amplify the decoy effects, leading to more pronounced differences in credibility ratings for target articles in the treatment groups.

\section{Discussion}
\subsection{Decoy Effect on Credibility Judgment}
Our study examined decoy effects on credibility assessments by both LLMs and human judges. Results from the two experiments showed that recent GPT and Claude models were generally effective at distinguishing credible information from misinformation, with correct articles receiving higher ratings across most COVID-19 treatment topics. Notably, LLMs, especially advanced models such as GPT-4o and Claude-3-Sonnet, were more susceptible to decoy effects compared to human judgers, particularly when using multiple-query or session-aware prompts that retain session context. In contrast, human judges exhibited decoy effects differently across varying topics, influenced by contextual features, such as prior knowledge and the sequence of article interaction. These findings emphasize the need to characterize and measure decoy effects in both human and AI-based information assessments.

Unlike supervised learning, LLMs can accumulate new knowledge by prompt input in a a natural language format without a training phase involved. This approach (in-context learning) allows LLMs to address news tasks without fine-turning the model. Our study found the in-context learning may not only accumulate new knowledge but also enhance bias potentially. We compared the single-query and multiple-query prompts in assessing information credibility. Multiple-query prompt, in which previous query would be retained as context for the current query, can enhance the decoy effects of articles with salient misinformation in the current query.

While social and algorithmic biases have been widely discussed in Machine Learning and AI fairness communities, how LLMs behave under cognitive bias triggers still remain understudied. Our study offers a unique perspective by comparing how advanced LLMs and human judges respond to decoy effect triggers in credibility assessments across varying search contexts. By examining the interaction between model sophistication and susceptibility to bias, our findings reveal that while newer models (e.g. GPT-4o and Claude-3-Sonnet) generally perform well in distinguishing credible information, they remain vulnerable to decoy effects, particularly in multi-query settings. This finding indicates that it is essential to consider not only accuracy and performance, but also the resilience to contextual biases in automated judgment tasks. Our research contributes to the ongoing discourse on AI reliability, harms, and risks by highlighting the urgent need for evaluation frameworks that account for the complexities of real-world information processing and potential cognitive biases, especially in domains where the stakes for accurate and unbiased judgments are high. These insights pave the way for more nuanced approaches to improving both human and AI-driven evaluations.

\subsection{Automated Judgment and AI Debiasing}
The increasing reliance on Generative AI for information judgment causes important concerns regarding cognitive biases embedded within these systems. LLMs, while adept at processing and generating text, are influenced by biases present in training data, which can lead to outputs that reflect and potentially reinforce misinformation and skewed interpretations. Examples include confirmation bias, where LLMs may prioritize information aligning with dominant viewpoints, and anchoring bias, where initial information disproportionately shapes subsequent judgments. Additionally, the decoy effect poses a specific risk, as LLMs might unintentionally emphasize certain information over others due to contextual presentation, leading to distorted credibility judgments. These risks are particularly critical in areas such as health and medical evaluation and decision-making, where accuracy and fairness are essential. 

To mitigate these risks, it is crucial to build and assess AI debiasing and auditing strategies. Debiasing could involve enhancing training datasets to better represent diverse perspectives and designing algorithms that detect and mitigate cognitive biases during the LLM's information synthesizing and judgment processes. The potential for the decoy effect to influence LLM outputs revealed in our study further highlights the need for rigorous auditing systems that monitor how information is presented and prioritized. This could include real-time auditing tools that flag potentially biased outputs and regular external audits of AI systems in sensitive domains. Collaboration between policymakers, researchers, and developers is essential to establish frameworks that ensure AI systems are transparent, accountable, and fair. Addressing cognitive biases in LLMs will contribute to more reliable and ethically sound AI-driven information environments across various fields. Our experiments underscore the heightened importance and urgency of implementing debiasing and auditing measures for larger and more advanced LLMs. As these models increase in complexity and potential vulnerability to bias triggers, the potential for cognitive biases to impact their outputs becomes more pronounced.

\subsection{Limitation and Future Directions}

Our study contributes valuable insights into the intersection of cognitive biases, credibility assessments, and AI systems, while also recognizing several limitations that present promising directions for future research. First, our experiments, which examined both crowdsourcing and large language model (LLM) systems, focused specifically on COVID-19 treatment information. While this focus allowed for an in-depth exploration of a highly relevant and critical area during the pandemic, it remains to be seen how the decoy effect on credibility judgments manifests across other domains. The decoy effect is a well-established cognitive bias in decision-making, and extending this research to domains such as legal, financial, and scientific information would offer a more comprehensive understanding of its broader impact.

Additionally, the LLMs used in our experiments were trained on different datasets with varying knowledge cut-off dates, much similar to the human assessors in crowdsourced studies who bring diverse backgrounds, expertise, and understanding of the judgment tasks. These variances suggest that different LLMs may exhibit differing levels of susceptibility to decoy manipulations. While this study did not directly compare these variations, the potential for LLMs trained on more up-to-date or specialized datasets to be more or less vulnerable to cognitive biases opens a rich avenue for future inquiry. Understanding these dynamics could significantly advance our understanding of how model training and data specificity affect AI performance in credibility judgments.

Despite these limitations, our work addresses a highly relevant and timely issue. COVID-19 treatment information has been critical for global populations, particularly marginalized communities who have been disproportionately impacted by both the pandemic and the spread of misinformation. By focusing on this pressing challenge, our study sheds light on how cognitive biases like the decoy effect can compromise the ability of both human and AI systems to accurately assess the credibility of health-related information. In this way, our research provides crucial insights into the vulnerabilities of AI systems in the context of medical misinformation, a global challenge that remains unresolved. These findings could guide the development of strategies to better protect users from the cognitive manipulations that contribute to the spread of misinformation. In addition, our experimental framework, designed to assess the decoy effect on both human and AI judgments, offers a replicable and scalable methodology. This framework can be adapted for use in different application domains, model types, user populations and task scenarios, making it a valuable tool and research resource for future studies, especially on the effectiveness and fairness of AI in judgment tasks. By applying this approach to other domains and exploring a broader range of tasks and user characteristics, researchers can deepen our understanding of how cognitive biases like the decoy effect influence decision-making across varied contexts.

Moreover, our work paves the way for future exploration of bias mitigation techniques that could be applied to LLMs at relatively low cost. Strategies such as prompt engineering and model fine-tuning hold promise for reducing AI systems' susceptibility to decoy effects and other cognitive biases, and future studies could evaluate the effectiveness of these interventions. Further, by extending our decoy effect designs, researchers could investigate the moderating effects of contextual factors, such as task complexity, question structures, or conversational interactions, on the strength of cognitive biases. These investigations would yield valuable insights into how to design AI systems that are more robust against biases, ultimately improving the trustworthiness and reliability of AI-driven information systems.

\section{Conclusion}
Can AI be \textit{cognitively biased}, especially in critical judgment tasks? In other words: If AI can (partially) do human information processing tasks, would they be "purely rational", or would they also make human-like mistakes under cognitive bias triggers? Motivated by this broad research problem, our study, focusing on the potential \textit{decoy effect} on both human and LLM judges, investigated the impact of decoy results on the credibility assessments on COVID medical (mis)information. In contrast to the generally assumed "rationality" of AI (especially by the general public who increasingly use AI in a variety of information retrieval, judgment and decision-making scenarios), our study empirically confirms the cognitive bias risks embedded in LLM agents, evaluates the decoy impact on LLMs against human assessors' behaviors, and highlights the complexity and importance of debiasing and auditing AI agents in judgment tasks. Our research reveals a new, cognitive bias dimension of AI bias identification and debiasing, which differs from widely discussed social and algorithmic biases, and can enhance our understanding of LLM's limitations in conducting and facilitating human judgment tasks. Our human-LLM dual experimental design can also inform future studies on exploring, evaluating, and mitigating LLM's cognitive biases of varying types (e.g. reference dependence~\cite{liu2020investigating, chen2023reference}, confirmation bias~\cite{liu2023behavioral, wang2024cognitively, white2013beliefs}, expectation disconfirmation~\cite{wang2024understanding}, threshold priming~\cite{scholer2013effect}) inherited from human-generated data, and facilitate the development of bias-aware LLM evaluation benchmarks in human-centered IR and AI research. 
\bibliographystyle{ACM-Reference-Format}
\bibliography{myref}

\end{document}